\documentclass[a4paper,11pt]{article}
\usepackage[utf8]{inputenc}
\usepackage[utf8]{inputenc}
\usepackage{bm}
\usepackage{amsmath}
\usepackage{amsfonts}
\usepackage{amssymb} 
\usepackage{graphicx}
\usepackage{listings} 
\usepackage{makeidx}
\makeindex

 \usepackage{float}
 \usepackage{caption,subcaption}
\usepackage{geometry}
\usepackage{color}

\usepackage{epstopdf}
\usepackage{sidecap}
\def\LW{\dimexpr.25\linewidth-.5em}

\begin{document}

\title{\bf Shallow Water Dynamics on Linear Shear Flows and Plane Beaches}
\author{Maria Bj\o rnestad\footnote{Department of Mathematics, University of Bergen,
        5020 Bergen, Norway, maria.bjoernestad@gmail.com,}~
 and Henrik Kalisch\footnote{Department of Mathematics, University of Bergen,
        5020 Bergen, Norway, henrik.kalisch@uib.no}}

\maketitle

\abstract{Long waves in shallow water propagating over a background shear flow 
towards a sloping beach are being investigated. 
The classical shallow-water equations are extended to incorporate both a background shear flow
and a linear beach profile, resulting in a non-reducible hyperbolic system. Nevertheless, it is shown
how several changes of variables based on the hodograph transform may be used to transform
the system into a linear equation which may be solved exactly using the method of separation
of variables. This method can be used to investigate the run-up of a long wave on a planar beach
including the development of the shoreline.}

\section{Introduction}
While many classical results in the theory of surface water waves 
have been obtained in the context of irrotational flow, the assumption 
of zero vorticity is not always justified.
Indeed, it is well known that vorticity may have a strong effect on the properties
of surface waves, and there is now a growing literature on the effect of vorticity on
the properties of surface waves. In mathematical studies focused on the influence of
vorticity on the dynamics of a free surface, some simplifying assumptions are usually made.
Examples of cases which have proved to be mathematically tractable include
compactly supported vorticity, such as point vortices or vortex patches \cite{CurtisKalisch, ShatahWalshZeng}, 
and the creation of vorticity through interaction with bathymetry \cite{CastroLannes} 
or through singular flow such as hydraulic jumps \cite{RG2013}.

One important case which is particularly amenable to both analytic and numerical methods
is the propagation of waves over a linear shear current. As noted in the
classical paper \cite{daSilva88}, there is a certain scale separation
between long surface waves and typical shear profiles which justifies
the assumption that the shear is unaffected by the wave motion to the
order of accuracy afforded by the model, and moreover, the precise 
profile of the shear flow may be approximated with a linear shear.

In the current work, we consider the case where a background shear current interacts with a sloping beach.
In particular, suppose the seabed is given by $h(x) = -\alpha x$ (see Figure 1),
and in addition a background shear flow $U(z) = \Gamma_0 + \Gamma_1 z$ is imposed. 
As shown in Appendix 1, for long surface waves, 
a set of shallow-water equations may be derived from first principles.
The system has the form
\begin{align}\label{system1}
\eta_t+ \Big(\Gamma_0 (h+\eta) +\frac{\Gamma_1}{2}(\eta^2-h^2)+u (h+\eta) \Big)_x=0 &, \\
\label{system2}
u_t+ \Big(us+\frac{s^2}{2}+\frac{u^2}{2}+g\eta \Big)_x = 0 &,
\end{align}
where $\eta(x,t)$ describes the deflection of the free surface
at a point $x$ and a time $t$, and $u(x,t)$ represents the horizontal fluid velocity.
The function $s(x)=\Gamma_0+\alpha\Gamma_1x$, 
and in particular the coefficient $\alpha \Gamma_1$
represent the strength of the interaction between the sea-bed and the shear.
Note that this system is hyperbolic, but the inclusion of non-trivial bathymetry
makes the system irreducible. Nevertheless it will be shown in the body of this paper
that it is possible to employ a hodograph transform which aids in the construction 
of exact solutions of the system, and in particular
allows us to make predictions of the development of the waterline.

\begin{figure}[t!]
\centering
\includegraphics[width=8cm]{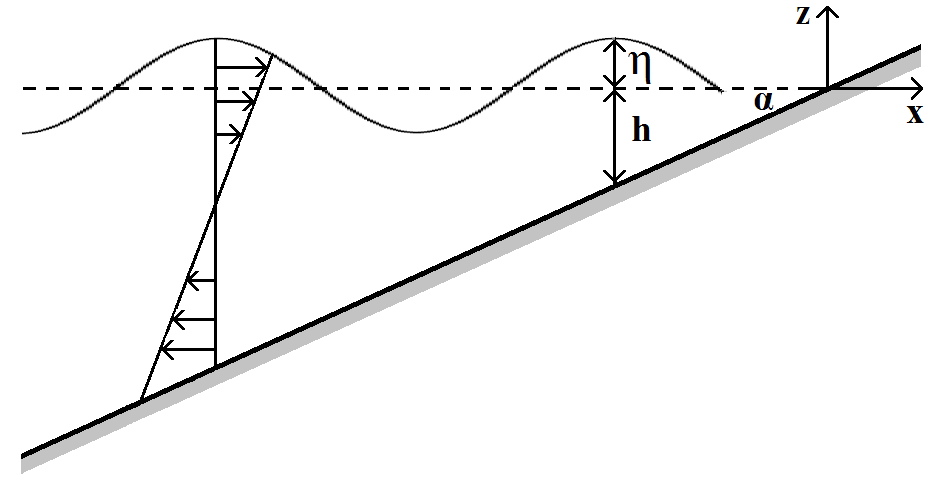}
\caption{\footnotesize Sloping beach given by $h(x)=-\alpha x$.}
\label{fig:slopebeachshear}
\end{figure}

The idea of exchanging the roles of dependent and independent variables
originated in the theory of gas dynamics \cite{CF}, and has been used in 
various special cases in hyperbolic equations, including the shallow-water equations.
However, it was not
until the work of Carrier and Greenspan \cite{CG} that it became
possible to find exact solutions for the shallow-water equations in the case of non-constant bathymetry.
Indeed, the real novelty of the work of Carrier and Greenspan lay in the fact that they succeeded in
applying the hodograph transform in the case of a non-uniform environment.
In particular, they obtained explicit solutions 
to the non-linear shallow-water equations on a linear beach profile, but without vorticity.

There are a few important variations on the method of Carrier and Greenspan.
In particular, more general initial data were considered in \cite{CWY}, and physical
properties such as mass and momentum fluxes related to the possible run-up of a tsunami
were mapped out. 
Some generalizations of the method with regards to the shape of the beach profile were made in \cite{DPS},
where a convex bottom topography of the type $h(x)= x^{4/3}$ was considered. Also, three-dimensional effects were
included in recent work \cite{RPD}, where a general approach was put forward to study
the problem on a bay of arbitrary cross-section. The work laid down in \cite{AntuonoBrocchini2}
makes use of analysis techniques to estimate the Jacobian function associated to an arbitrary
bottom profile, and thus proves that at least in theory, that the restriction to planar 
or convex beaches is not necessary.
One problematic issue with the approach of Carrier and Greenspan is that it is difficult to
treat the boundary-value problem. For example, if wave and velocity data are known at a fixed location
it is not straightforward to prescribe these as boundary data, and study the shoaling and run-up of the
resulting shorewards propagating waves. This problem was investigated in-depth in \cite{AntuonoBrocchini1},
where it was shown how the boundary-value can be solved in the context of planar beaches.

As we stated above, the main purpose of the current work is to extend the
Carrier-Greenspan approach to the case where background vorticity can be included
in the flow. The need for such an extension arises from the fact that
the propagation of water waves in coastal areas is often affected by the influence of currents. 
Previous works on this topic include the construction of periodic traveling waves over shear
flows in the Euler equations \cite{CV}, numerical investigations \cite{Vanden2}
and the investigation of the pressure profile in asymptotic models \cite{AK3,VO}.

The plan of the current paper is as follows. In section 2 we consider the case of a shear flow
over a flat bed. While the inclusion of background vorticity into shallow-water
models is known (see \cite{KharifAbid} for instance), it is not obvious
how to find closed-form expressions for the Riemann invariants in this case. 
In section 3, we treat the case of a shear flow over a linear beach,
and use intuition gained from the Riemann invariants in the flat-bed case
to aid in the construction of the hodograph transform in the more difficult
case of non-constant bathymetry.
Finally, in Section 4, we explain how the equations may be solved exactly,
and we include a few plots where we compare cases with different
strengths of background vorticity.
Finally, the equations with both shear flow and an uneven bottom are derived
in the Appendix.

\section{Shear flow over a flat bed}
\begin{figure}[h!]
\centering					
\includegraphics[width=10cm]{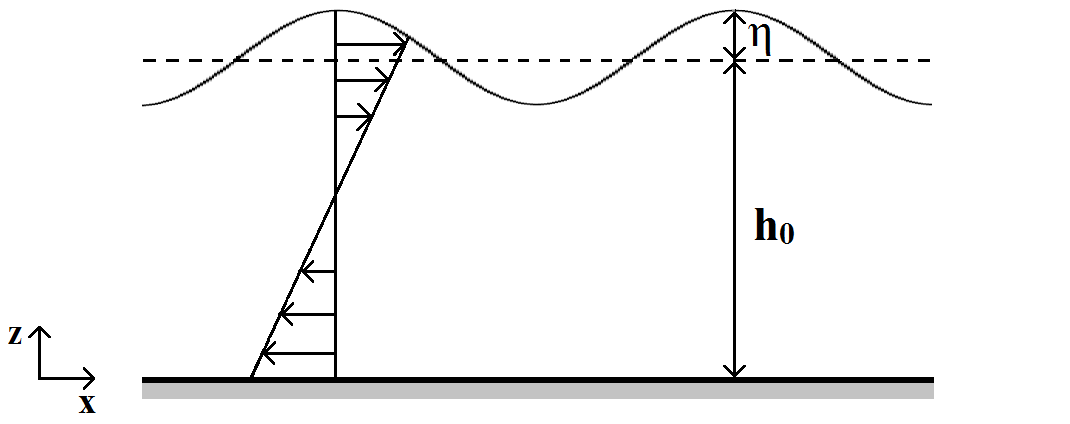}
\caption{\footnotesize Background shear flow for constant depth.}
\label{fig:flatshear}
\end{figure}

We first look at the case of shear flow over a flat bed as this case
will give us important clues on how to proceed in the more difficult
case of a shear flow over a sloping bed.
A sketch of the geometry is shown in Fig. \ref{fig:flatshear}. 
In particular, the total depth is $H(x,t) = \eta(x,t) +h_0$, 
where $h_0$ is the constant undisturbed depth.
The vertical shear current is assumed to be of the form $U(z)=-\Gamma_0+\Gamma_1 z$ 
which yields a background vorticity $-\Gamma_1$.
Without loss of generality, we may assume that the density is constant,
and consider a domain of unit width in the transverse direction.
The shallow-water equations for a flat bed are as follows:
\begin{equation}\label{flatbunnmass}
H_t + \Big( -\Gamma_0 H+\frac{\Gamma_1}{2}H^2  +  uH \Big)_x  = 0,
\end{equation}
\begin{equation}\label{flatbunnmomentum}
u_t + \Big( -\Gamma_0u +\frac{1}{2}u^2 + gH \Big)_x = 0.
\end{equation}
In order to express the equations in non-dimensional variables, we 
introduce the following scaling:  
$u^* = \frac{u}{u_0} \:,\:\:\: \eta^* = \frac{\eta}{h_0} \:,\:\:\: x^*=\frac{x}{h_0} \:,\:\:\: t^*=\frac{t}{T} \:,\:\:\: \Gamma_0^*=\frac{\Gamma_0}{u_0}\:,\:\:\: \Gamma_1^*=\frac{\Gamma_1}{1/ T}$ where $T=\sqrt{h_0/ g} \:,\:u_0=\sqrt{gh_0}\:$.  
The equations are the written in non-dimensional form as
\begin{equation}\nonumber
H^*_{t^*} + \Big( -\Gamma_0^* H^*+\frac{\Gamma_1^*}{2}H^{*2}  +  u^*H^* \Big)_{x^*}  = 0,
\end{equation}
\begin{equation}\nonumber
u^*_{t^*} + \Big(-\Gamma_0^*u^* +\frac{1}{2}u^{*2} + H^* \Big)_{x^*} = 0.
\end{equation}

As is customary in shallow-water theory, the propagating speed of a wave is taken as $c=\sqrt{gH}$ 
(in non-dimensional variables $c^*=\sqrt{H^*}$ where $c^*=\frac{c}{u_0}$). 
Note that for easier reading, the stars on the non-dimensional variables will be omitted from now on.
Adding and subtracting the two equations above, and
using the speed $c$ as an unknown, the equations can be written in so-called pre-characteristic form as
\begin{align*}
\left\lbrace\frac{\partial}{\partial t} + (u-\Gamma_0+c)\frac{\partial}{\partial x}\right\rbrace \left(u+2c\right) &= -2\Gamma_1c^2c_x, \\
\left\lbrace\frac{\partial}{\partial t} + (u-\Gamma_0-c)\frac{\partial}{\partial x}\right\rbrace \left(u-2c\right) &= 2\Gamma_1c^2c_x.
\end{align*}
This form may be useful in some situations connected to numerical integration of the equations,
but is included here mainly as a stepping stone toward a similar set of equations in the
case of the sloping bottom.
In the current context, it is actually more advantageous to put the equations 
into proper characteristic form.
However, since it is not easy to see how to eschew the $2\Gamma_1c^2c_x$-terms on the right hand side,
we will use a different approach to put the equations in characteristic form.

In vector notation, we can write eq.(\ref{flatbunnmass}) and eq.(\ref{flatbunnmomentum}) as 
\begin{equation}\label{flatbunnmatriseligning}
 \mathbf{u}_t+\mathbf{f}(\mathbf{u})_x=\mathbf{0}
\end{equation}
where $\mathbf{u}=\left[H,u\right]^T$. 
Further, $\mathbf{f}(\mathbf{u})_x=\mathbf{f}'(\mathbf{u})\mathbf{u}_x$, 
where $\mathbf{f}'(\mathbf{u})$ is the Jacobian matrix
\begin{equation} \nonumber
\mathbf{f}'(\mathbf{u})=
\begin{bmatrix}
-\Gamma_0+\Gamma_1H+u & H \\ 
1 & -\Gamma_0+u
\end{bmatrix}.
\end{equation}
The eigenvalues are
\begin{equation}\nonumber
\xi_1=u-\Gamma_0+\frac{1}{2}\Gamma_1H+\frac{1}{2}\sqrt{(\Gamma_1H)^2+4H},
\end{equation}
\begin{equation}\nonumber
\xi_2=u-\Gamma_0+\frac{1}{2}\Gamma_1H-\frac{1}{2}\sqrt{(\Gamma_1H)^2+4H}.
\end{equation}
These eigenvalues are real and distinct which means that the system is strictly hyperbolic. 
Since the Jacobian matrix only depends on $\mathbf{u}$, and not $x$ or $t$,
the system is reducible, and Riemann invariants exist according to the standard theory \cite{CF}.
However, finding exact expressions for the Riemann invariants is in general highly non-trivial.

In order to find the the Riemann invariants $\omega_1$ and $\omega_2$, 
it will be convenient to 
define an eigenproblem $\mathbf{L}\mathbf{f}'(\mathbf{u})=\mathbf{\Lambda}\mathbf{L}$ 
with the left eigenvectors 
\begin{equation} \label{flatbunnvenstreegenvector} \nonumber
\mathbf{l_1}=
\begin{bmatrix}
 2  \\  -\Gamma_1H + \sqrt{(\Gamma_1H)^2+4H}
\end{bmatrix},
\end{equation}
\begin{equation}\nonumber
\mathbf{l_2}=
\begin{bmatrix}
 2  \\  -\Gamma_1H - \sqrt{(\Gamma_1H)^2+4H}
\end{bmatrix}.
\end{equation}
Inserting the left eigenproblem in eq.(\ref{flatbunnmatriseligning}), 
we can express eq.(\ref{flatbunnmatriseligning}) as
\begin{equation}\label{indexformleftvectoreq}
\mathbf{l}_i^T\mathbf{u}_t+\xi_i\mathbf{l}_i^T\mathbf{u}_x=\mathbf{0},
\end{equation}
where $i=1,2$.
If we now introduce the auxiliary function $\mu(\mathbf{u})$ satisfying 
\begin{equation} \label{flatbunngradientomega}
\nabla\omega_i(\mathbf{u}) =
\begin{bmatrix}
 \frac{\partial\omega_i}{\partial H} \:,& \frac{\partial\omega_i}{\partial u}
\end{bmatrix}
=\mu_i(\mathbf{u}) \mathbf{l}_i^T,
\end{equation} 
the eq.(\ref{indexformleftvectoreq}) can be written as
\begin{equation}
\nabla\omega_i(\mathbf{u})\mathbf{u}_t+\xi_i \nabla\omega_i(\mathbf{u})\mathbf{u}_x=0,
\end{equation}
which is the same as
\begin{equation}\nonumber
\left\lbrace \frac{\partial}{\partial t}+\xi_i\frac{\partial}{\partial x}\right\rbrace\omega_i(\mathbf{u}) =0.
\end{equation} 
The characteristic form in the latter equation shows that $\omega_i(\mathbf{u})$ is constant along the characteristics $\frac{dx}{dt}=\xi_i(\mathbf{u})$. 
The challenging part of this procedure is to find an expression for $\mu_i(\mathbf{u})$. 
To be able to proceed further, we start by assuming that $\mu_i(\mathbf{u})$
is chosen such that the relation
$\frac{\partial^2\omega_i}{\partial H\partial u}=\frac{\partial^2\omega_i}{\partial u\partial H}$
is satisfied.
First, to calculate $\mu_1(\mathbf{u})$, eq.(\ref{flatbunngradientomega}) gives us
\begin{align} \label{flatbunnmixderiverte}
\frac{\partial\omega_1}{\partial H} &= 2 \mu_1(\mathbf{u}) \nonumber \\
\frac{\partial\omega_1}{\partial u} &= \mu_1(\mathbf{u}) \Big(-\Gamma_1H + \sqrt{(\Gamma_1H)^2+4H} \Big),
\end{align}
and if we let 
\begin{equation}
\mu_1(\mathbf{u})= \Gamma_1+\frac{1}{H}\sqrt{(\Gamma_1H)^2+4H},
\end{equation}
the assumption will be satisfied. Integration gives us
\begin{align}
\omega_1 &= 2\Gamma_1H+2\sqrt{(\Gamma_1H)^2+4H}+\frac{8}{\Gamma_1}\sinh^{-1}\left(\frac{\Gamma_1\sqrt{H}}{2}\right)+ K_1(u),\nonumber \\
\omega_1 &= 4u + K_2(H),\nonumber
\end{align}
where $K_1(H)$ and $K_2(u)$ are the constants of integration. By combining these, we obtain the first Riemann invariant 
\begin{equation}
\omega_1 = u -\Gamma_0+  \frac{1}{2}\Gamma_1H+\frac{1}{2}\sqrt{(\Gamma_1H)^2+4H}+\frac{2}{\Gamma_1}\sinh^{-1}\left(\frac{\Gamma_1\sqrt{H}}{2}\right),
\end{equation}
where we also have divided by 4 and subtracted by $\Gamma_0$ to simplify further work.

We can obtain the second Riemann invariant in a similar way.
With the expression for the parameter $\mu_2$ given by
\begin{equation*}
\mu_2(\mathbf{u})= \Gamma_1-\frac{1}{H}\sqrt{(\Gamma_1H)^2+4H},
\end{equation*}
we get
\begin{equation*}
\omega_2 = u-\Gamma_0 +  \frac{1}{2}\Gamma_1H-\frac{1}{2}\sqrt{(\Gamma_1H)^2+4H}-\frac{2}{\Gamma_1}\sinh^{-1}\left(\frac{\Gamma_1\sqrt{H}}{2}\right).
\end{equation*}
With these expressions in hand, the equations (\ref{flatbunnmass}) and (\ref{flatbunnmomentum}) 
can then be rewritten in characteristic form as
\begin{align*}
\left\lbrace \frac{\partial}{\partial t}+\xi_1\frac{\partial}{\partial x}\right\rbrace\omega_1 &=0, \\ \left\lbrace \frac{\partial}{\partial t}+\xi_2\frac{\partial}{\partial x}\right\rbrace\omega_2 &=0.
\end{align*}
However, the equations are still nonlinear.
Since one purpose of the present study is to obtain exact representations of solutions
of (\ref{flatbunnmass}) and (\ref{flatbunnmomentum}), it will be convenient to perform
yet another transformation to put the equations in linear form. 
 
Switching dependent and independent variables via a hodograph transform 
from $\omega_1=\omega_1(x,t)$ and $\omega_2=\omega_2(x,t)$ to $x=x(\omega_1,\omega_2)$ and $t=t(\omega_1,\omega_2)$, 
results in
\begin{align} \label{hodo flat en}
x_{\omega_2}-\xi_1 t_{\omega_2}&=0, \\
x_{\omega_1}-\xi_2 t_{\omega_1}&=0. \label{hodo flat to}
\end{align}
As long as the Jacobian matrix remains non-singular, linearity has been achieved 
and the equations can now be solved. We will come back to the solution in section 4.1.
%\newpage

\section{Shear flow on a sloping bed}
We will now consider the geometry in Fig. \ref{fig:slopebeachshear} 
with the total depth  $H(x,t)=\eta(x,t)+h(x)$.
The vertical shear current is assumed to be of the form $U(z)=\Gamma_0+\Gamma_1 z$ 
with the vorticity $-\Gamma_1$.
Note that the $x$-axis is now assumed to be aligned with the undisturbed
free surface as this normalization is more convenient in the current setting.

To put equations (\ref{system1}) and (\ref{system2}) into non-dimensional form, 
we introduce new variables $u^* = \frac{u}{u_0} \:,\:\:\: \eta^* = \frac{\eta}{\alpha l_0} \:,\:\:\: x^*=\frac{x}{l_0}\:,\:\:\: t^*=\frac{t}{T} \:,\:\:\: \Gamma_0^*=\frac{\Gamma_0}{u_0}\:,\:\:\: \Gamma_1^*=\frac{\Gamma_1}{1/ T}$ 
where $T=\sqrt{l_0/\alpha g} \:,\:\:\: u_0=\sqrt{gl_0\alpha}\:$  and $l_0$ is a characteristic length.
We also define $s^*=\frac{s}{u_0}$.
The equations then appear as
\begin{equation}\label{nondim mass slope}
\eta^*_{t^*}+ \Big(\Gamma_0^*(\eta^*-x^*)+\frac{\alpha\Gamma_1^*}{2}(\eta^{*2}-x^{*2})+u^*(\eta^*-x^*)\Big)_{x^*}=0,
\end{equation}
\begin{equation}\label{nondim mom slope}
u^*_{t^*}+ \Big(u^*s^*+\frac{s^{*2}}{2}+\frac{u^{*2}}{2}+\eta^* \Big)_{x^*} = 0.
\end{equation}
As in the previous section for the sake of readability, the stars will be disregarded in what follows.
In an attempt to write the equations in characteristic form,
one may insert the propagation speed in non-dimensional form $c=\sqrt{(\eta-x)}$, 
and then add and subtract them to obtain 
the pre-characteristic form 
\begin{align*}
\left\lbrace\frac{\partial}{\partial t} + (u+s+c)\frac{\partial}{\partial x}\right\rbrace \left(u+s+2c+t\right) &= -2s_xc^2c_x, \\
\left\lbrace\frac{\partial}{\partial t} + (u+s-c)\frac{\partial}{\partial x}\right\rbrace \left(u+s-2c+t\right) &= 2s_xc^2c_x.
\end{align*}
To be able to solve these equations, the difficulty lies in finding the Riemann invariants.
We can write eq.(\ref{nondim mass slope}) and eq.(\ref{nondim mom slope}) 
as $\mathbf{u}_t+\mathbf{f}(\mathbf{u},x)_x=\mathbf{0}$ where $\mathbf{u}=\left[\eta,u\right]^T$. 
The Jacobian matrix $\mathbf{f}'(\mathbf{u},x)$ has the following eigenvalues
\begin{align}
\xi_1&=u+s+\frac{\alpha\Gamma_1}{2}c^2+\frac{c}{2}\sqrt{(\alpha\Gamma_1c)^2+4},\nonumber \\
\xi_2&=u+s+\frac{\alpha\Gamma_1}{2}c^2-\frac{c}{2}\sqrt{(\alpha\Gamma_1c)^2+4}. \nonumber
\end{align}
Since the Jacobian matrix now depends on $x$, the system is not reducible,
and it is not clear whether Riemann invariants can be found.
In particular we cannot proceed in the same way as in section 2. 
However, when carefully combining the pre-characteristic form and the eigenvalues 
with the corresponding equations for the flat bed case,
a bit of informed guessing  points to defining the Riemann invariants as
\begin{align*}
\omega_1 &= u +s +  \frac{1}{2}\alpha\Gamma_1c^2+\frac{1}{2}c\sqrt{(\alpha\Gamma_1c)^2+4}+\frac{2}{\alpha\Gamma_1}\sinh^{-1}\left(\frac{\alpha\Gamma_1c}{2}\right) +t \\
\omega_2 &= u +s + \frac{1}{2}\alpha\Gamma_1c^2-\frac{1}{2}c\sqrt{(\alpha\Gamma_1c)^2+4}-\frac{2}{\alpha\Gamma_1}\sinh^{-1}\left(\frac{\alpha\Gamma_1c}{2}\right) +t
\end{align*}
As it turns out, if these expressions are substituted into eq.(\ref{nondim mass slope}) 
and eq.(\ref{nondim mom slope}), the characteristic form
\begin{equation*}
\left\lbrace\frac{\partial}{\partial t} + \xi_1\frac{\partial}{\partial x}\right\rbrace \omega_1 =0,
\end{equation*}
\begin{equation*}
\left\lbrace\frac{\partial}{\partial t} + \xi_2\frac{\partial}{\partial x}\right\rbrace \omega_2 =0
\end{equation*}
appears. These two equations are still nonlinear in $t$, so we continue by performing 
a hodograph transformation, changing $\omega_1=\omega_1(x,t)$ and $\omega_2=\omega_2(x,t)$ 
to  $x=x(\omega_1,\omega_2)$ and $t=t(\omega_1,\omega_2)$, which results in the equations
\begin{align*} 
x_{\omega_2}-\xi_1 t_{\omega_2}&=0, \\
x_{\omega_1}-\xi_2 t_{\omega_1}&=0.
\end{align*}
In contrast to the flat bed case, the equations are still nonlinear at this stage. 
Therefore, another step is required, 
and we introduce new variables $\omega_1+\omega_2=\lambda$ and $\omega_1-\omega_2=\sigma$. 
This change of variables give us
\begin{align} \label{x_lambda-equation}
x_\lambda-At_\lambda+Bt_\sigma&=0, \\ \label{x_sigma-equation}
x_\sigma-At_\sigma+Bt_\lambda&=0,
\end{align}
where to simplify, we have defined $A=u+s+\frac{\alpha\Gamma_1}{2}c^2$ 
and $B=\frac{c}{2}\sqrt{(\alpha\Gamma_1c)^2+4}$. 
Further, differentiating these equations, and using 
the identities $x_{\sigma\lambda}=x_{\lambda\sigma}$ and $t_{\sigma\lambda}=t_{\lambda\sigma}$
leads to 
\begin{equation} \label{AogBligning}
A_\lambda t_\sigma-A_\sigma t_\lambda-B_\lambda t_\lambda+B_\sigma t_\sigma = B\left(t_{\lambda\lambda}-t_{\sigma\sigma} \right).
\end{equation}
In order to find expressions for the derivatives of $A$ and $B$ 
with respect to $\sigma$ and $\lambda$, we start by writing the variables $\lambda$ and $\sigma$ as
\begin{align} \label{lambda/2}
\frac{\lambda}{2} &=u+s+\frac{\alpha\Gamma_1}{2}c^2+t, \\ \label{sigma/2}
\frac{\sigma}{2} &=\frac{c}{2}\sqrt{(\alpha\Gamma_1c)^2+4}+\frac{2}{\alpha\Gamma_1}\sinh^{-1}\left(\frac{\alpha\Gamma_1c}{2}\right).
\end{align}
It is easy to see from eq.(\ref{lambda/2}) that $A_\sigma=-t_\sigma$ and $A_\lambda=\frac{1}{2}-t_\lambda$.
To calculate $B_\lambda$ and $B_\sigma$, we start by differentiating $B$ to find
\begin{equation*}
B_\sigma=\frac{(\alpha\Gamma_1c)^2+2}{\sqrt{(\alpha\Gamma_1c)^2+4}}c_\sigma ,\:\:\:\:\:
B_\lambda=\frac{(\alpha\Gamma_1c)^2+2}{\sqrt{(\alpha\Gamma_1c)^2+4}}c_\lambda, 
\end{equation*}
where $c_\sigma$ and $c_\lambda$ are unknown.  We can find an expression for these 
by differentiating eq.(\ref{sigma/2}) implicitly with respect to $\sigma$ and $\lambda$, yielding
\begin{equation} \label{c_sigma}
\frac{1}{2}=c_\sigma \sqrt{(\alpha\Gamma_1c)^2+4} ,\:\:\:\:\: 0=c_\lambda \sqrt{(\alpha\Gamma_1c)^2+4}.
\end{equation}
Since the root cannot be zero, $c_\lambda$ has to be zero. Thus, with these calculations eq.(\ref{AogBligning}) becomes
\begin{equation}\label{mellom}
\left(\frac{(\alpha\Gamma_1c)^2+3}{(\alpha\Gamma_1c)^2+4}\right)t_\sigma=\frac{c}{2}\sqrt{(\alpha\Gamma_1c)^2+4}\left(t_{\lambda\lambda}-t_{\sigma\sigma}\right).
\end{equation}
Unfortunately, the $c$ is only given implicitly as a function of $\sigma$ in eq.(\ref{sigma/2}). 
However, notice that in eq.(\ref{sigma/2}) both terms are increasing and monotone, 
so the relation can be inverted. 
Since we seek an expression for $c_{\sigma\sigma}$, we start by differentiating eq.(\ref{sigma/2}) twice and get
\begin{equation}\nonumber
0=c_{\sigma\sigma}\sqrt{(\alpha\Gamma_1c)^2+4}+\frac{(\alpha\Gamma_1c_\sigma)^2c}{\sqrt{(\alpha\Gamma_1c)^2+4}}.
\end{equation}
By inserting $c_\sigma$ from eq.(\ref{c_sigma}), we obtain the expression
\begin{equation}\nonumber
c_{\sigma\sigma}=-\frac{(\alpha\Gamma_1)^2c}{4((\alpha\Gamma_1c)^2+4)^2}.
\end{equation}
With some calculations eq.(\ref{mellom}) then becomes
\begin{equation}\label{linear t_(c,lambda)-equation}
ct_{cc}+3t_c=4c((\alpha\Gamma_1c)^2+4)t_{\lambda\lambda},
\end{equation}
which is a linear equation and can now be solved exactly.

\section{Exact solutions of the equations}
\subsection{Flat bed}
One way to solve eq.(\ref{hodo flat en}) and eq.(\ref{hodo flat to}) is to
introduce new variables in the same way as shown above for the
case of the sloping bed.
Thus, introducing the variables $\lambda = \omega_1+\omega_2$ and $\sigma = \omega_1-\omega_2$, 
the equations can be written as
\begin{align} \label{x_lambdaflat-equation}
x_\lambda-(u-\Gamma_0+\frac{1}{2}\Gamma_1 H)t_\lambda+\frac{1}{2}\sqrt{(\Gamma_1H)^2+4H}\:t_\sigma&=0, \\ \label{x_sigmaflat-equation}
x_\sigma-(u-\Gamma_0+\frac{1}{2}\Gamma_1 H)t_\sigma+\frac{1}{2}\sqrt{(\Gamma_1H)^2+4H}\:t_\lambda&=0.
\end{align} 
Moreover remembering the expressions for $\omega_1$ and $\omega_2$ from Section 2,
$\lambda$ and $\sigma$ appear as
\begin{align} \label{flatlambda/2}
\frac{\lambda}{2} &=u-\Gamma_0+\frac{\Gamma_1}{2}H, \\ \label{flatsigma/2}
\frac{\sigma}{2} &=\frac{1}{2}\sqrt{(\Gamma_1H)^2+4H}+\frac{2}{\Gamma_1}\sinh^{-1}\left(\frac{\Gamma_1\sqrt{H}}{2}\right).
\end{align}
Inverting the relation (\ref{flatsigma/2}) results in the following linear equation for $t(H,\lambda)$:
\begin{equation}\label{flatlinear t(c,lambda)-equation}
Ht_{HH}+2t_H=(\Gamma_1^2H+4)t_{\lambda\lambda}.
\end{equation}
Before we solve this equation, 
notice that it is problematic to calculate $x(H,\lambda)$ 
without introducing a 'potential' function for $t(H,\lambda)$, i.e.
\begin{equation}\label{flat t potential}
t=\frac{1}{\Gamma_1^2H+4}\:\phi_H.
\end{equation}
However, if this potential is used, eq.(\ref{x_sigmaflat-equation}) 
gives us an expression for $x(H,\lambda)$, viz.
\begin{equation*}
x=\frac{\lambda}{2}t-\frac{1}{2}\phi_\lambda.
\end{equation*}
Eq.(\ref{flatlinear t(c,lambda)-equation}) can now be written due to eq.(\ref{flat t potential}) as
\begin{equation*}
H(\Gamma_1^2H+4)\phi_{HH}+4\phi_H = \left(\Gamma_1^2H+4\right)\phi_{\lambda\lambda}
\end{equation*}
This equation can be solved using separation of variables, and the solution has the general form
\begin{equation*}
\phi(H,\lambda)=A\cos(\omega\lambda)e^{-i\Gamma_1\omega H}\left[-\omega H(i\Gamma_1-2\omega) \mathcal{F}_1+(i\Gamma_1\omega H-1)\mathcal{F}_2\right]
\end{equation*}
where 
\begin{equation}\nonumber
\mathcal{F}_1= \, _{1} \! \, F_{1}\left(\frac{2i\omega+2\Gamma_1}{\Gamma_1},3,2i\Gamma_1\omega H\right),\:\:\:\:\: 
\mathcal{F}_2= \, _{1} \! \, F_{1}\left(\frac{2i\omega+\Gamma_1}{\Gamma_1},2,2i\Gamma_1\omega H\right)
\end{equation}
are given in terms the generalized hypergeometric function $_{1}F_{1}$ \cite{OlverDLMF}.
Finally, the principal unknowns can be expressed in terms of $\lambda$ and $H$ as
$
u= \frac{\lambda}{2}+\Gamma_0-\frac{\Gamma_1}{2}H
$ 
and $\eta=H-h_0$.

\subsection{Sloping bed}
We now look at the more interesting case of exact solutions in the presence
of the inclined bottom profile.
To be able to solve for $x(c,\lambda)$, we will also here make use of a 'potential' function.
Instead of introducing the potential function for $t(c,\lambda)$ directly, we rather start by defining
\begin{equation}\label{W_c,lambda}
W(c,\lambda)=u(c,\lambda)+\alpha\Gamma_1x(c,\lambda)+\frac{\alpha\Gamma_1}{2}c^2.
\end{equation}
Combining the new function $W(c,\lambda)$ with eq.(\ref{lambda/2}), we can rewrite eq.(\ref{linear t_(c,lambda)-equation}) and obtain
\begin{equation}\label{slope W equation}
cW_{cc}+3W_c=4c\left((\alpha\Gamma_1c)^2+4\right) W_{\lambda\lambda}.
\end{equation}
If we now define the function $\phi(c,\lambda)$ by
\begin{equation}\label{slopeW}
W(c,\lambda)=\frac{1}{c\left((\alpha\Gamma_1c)^2+4\right)}\:\phi_c(c,\lambda),
\end{equation}
then eq.(\ref{slope W equation}) becomes
\begin{equation}\label{pde}
c \phi_{cc}+\frac{4-(\alpha \Gamma_1c)^2}{4+(\alpha\Gamma_1c)^2}\phi_c = 4c\left((\alpha\Gamma_1 c)^2+4\right)\phi_{\lambda\lambda}.
\end{equation}
We seek a solution in the form $\phi(c,\lambda)=f(c)g(\lambda)$, and thus separating the variables gives 
two equations of the form
\begin{equation}\nonumber
c \left( (\alpha\Gamma_1 c)^2+4 \right) f''(c)+ \left(4- (\alpha \Gamma_1c)^2\right) f'(c)+4\omega^2c \left((\alpha\Gamma_1 c)^2+4\right)^2 f(c)=0,
\end{equation}
\begin{equation}\nonumber
g''(\lambda)+\omega^2 g(\lambda)=0
\end{equation}
where $\omega$ is a constant. The solution $\phi(c,\lambda)$ should be bounded as $c\rightarrow 0$, 
and the corresponding solution of (\ref{pde}) is 
\begin{equation}\nonumber
\phi(c,\lambda)=A\cos(\omega\lambda)e^{-i\alpha\Gamma_1\omega c^2}\left[-\omega c^2(i\alpha\Gamma_1-2\omega) 
\mathcal{F}_1+(i\alpha\Gamma_1\omega c^2-1)\mathcal{F}_2\right],
\end{equation}
where $\mathcal{F}_1$ and $\mathcal{F}_2$ are defined in terms of the 
generalized hypergeometric functions $_1\!F_{1}$, evaluated with the following arguments:
\begin{equation}\nonumber
\mathcal{F}_1 = _1\!\!F_{1}\left(\frac{2i\omega+2\alpha\Gamma_1}{\alpha\Gamma_1},3,2i\alpha\Gamma_1\omega c^2\right),\:\:\:\:\: \mathcal{F}_2= _1\!\!F_{1}\left(\frac{2i\omega+\alpha\Gamma_1}{\alpha\Gamma_1},2,2i\alpha\Gamma_1\omega c^2\right).
\end{equation}
Using the function $W(c,\lambda)$, an expression for $t(c,\lambda)$ 
can be obtained from eq.(\ref{lambda/2}):
\begin{equation}\nonumber
t=\frac{\lambda}{2}-W-\Gamma_0. 
\end{equation}
Further, an expression for $x(c,\lambda)$ can be obtained from eq.(\ref{x_sigma-equation}). 
Inserting for $t(c,\lambda)$ from eq.(\ref{lambda/2}) and eq.(\ref{W_c,lambda}), results in
\begin{equation}\nonumber
x_c=-WW_c-\Gamma_0 W_c-c\left((\alpha\Gamma_1 c)^2+4\right)\left(\frac{1}{2}-W_\lambda\right), 
\end{equation}
and in terms of the function $\phi$, it becomes
\begin{equation}\nonumber
x=-\frac{W^2}{2}-\Gamma_0 W-\frac{c^2}{8}\left((\alpha\Gamma_1 c)^2+8\right) +\phi_\lambda.
\end{equation}
The equation for the propagation speed gives us the free surface elevation as $\eta (c,\lambda)=c^2+x(c,\lambda)$,
and an expression for the velocity component $u(c,\lambda)$ is given by
\begin{equation}\nonumber
u=W-\alpha\Gamma_1x-\frac{\alpha\Gamma_1}{2}c^2
\end{equation}
from eq.(\ref{lambda/2}) and eq.(\ref{W_c,lambda}).
\begin{figure}[b!]
\begin{center}
\parbox{\LW}{\includegraphics[width=2.1\linewidth]{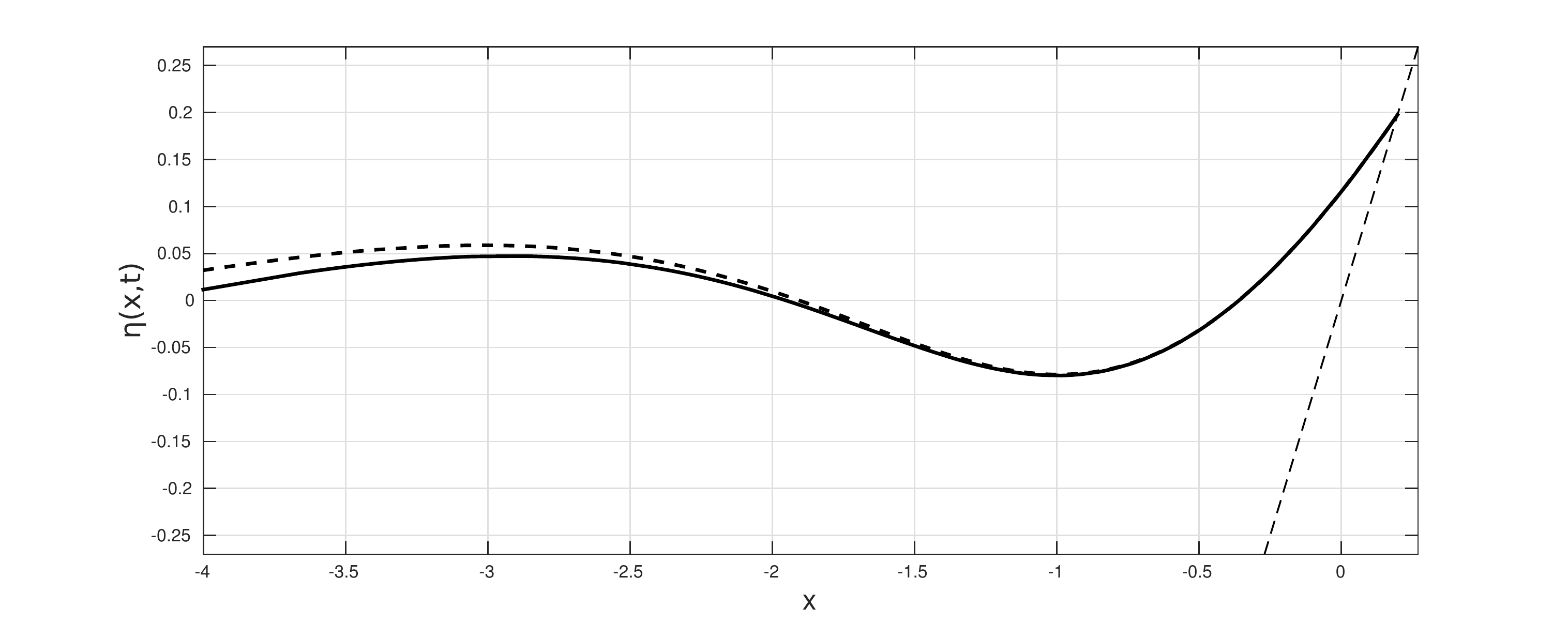}}\hfill%
\parbox{\LW}{$t_1=0.89$}\hfill \\
\parbox{\LW}{\includegraphics[width=2.1\linewidth]{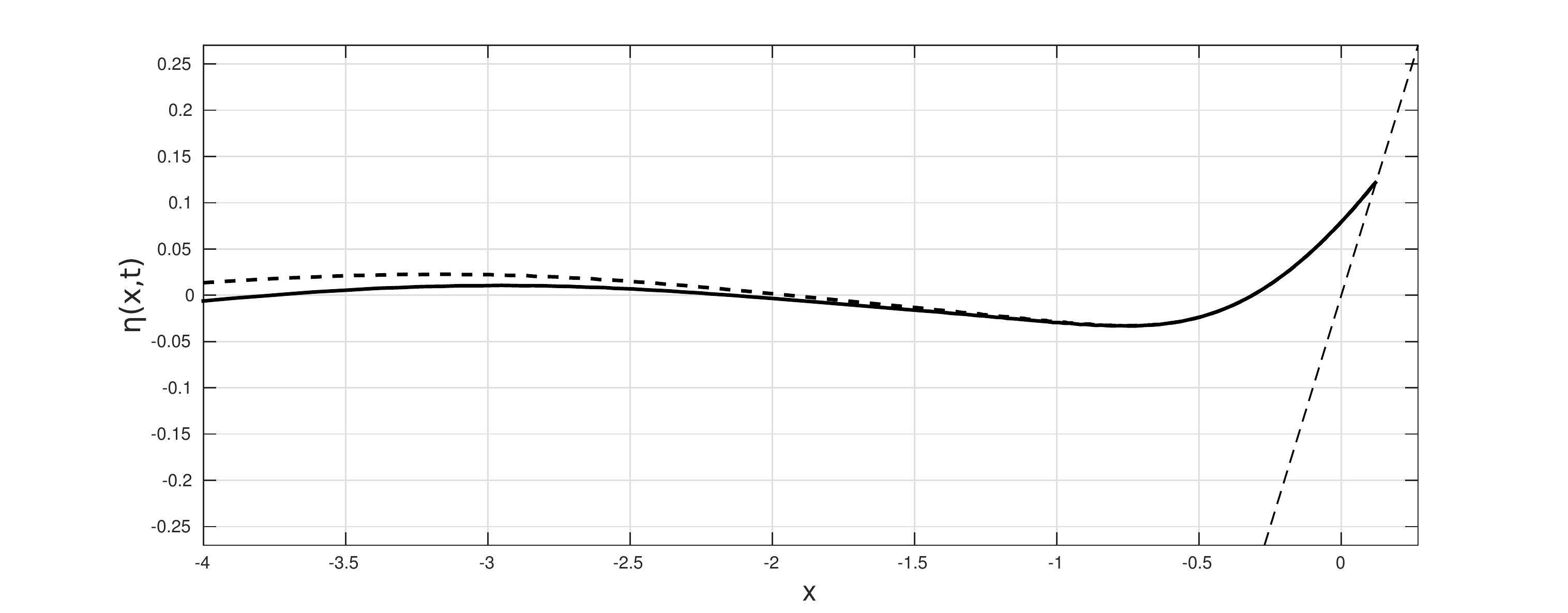}}\hfill%
\parbox{\LW}{$t_2=1.38$}\\
\parbox{\LW}{\includegraphics[width=2.1\linewidth]{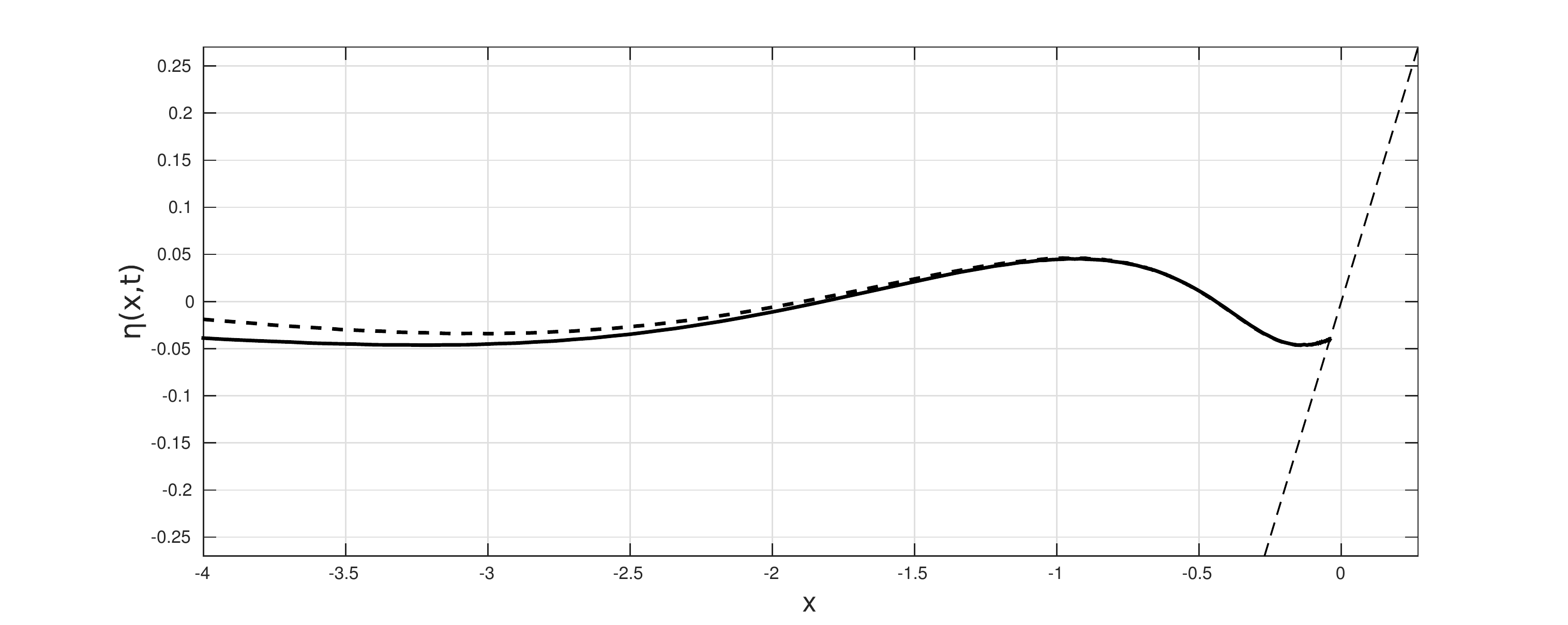}}\hfill%
\parbox{\LW}{$t_3=1.87$}\hfill \\
\parbox{\LW}{\includegraphics[width=2.1\linewidth]{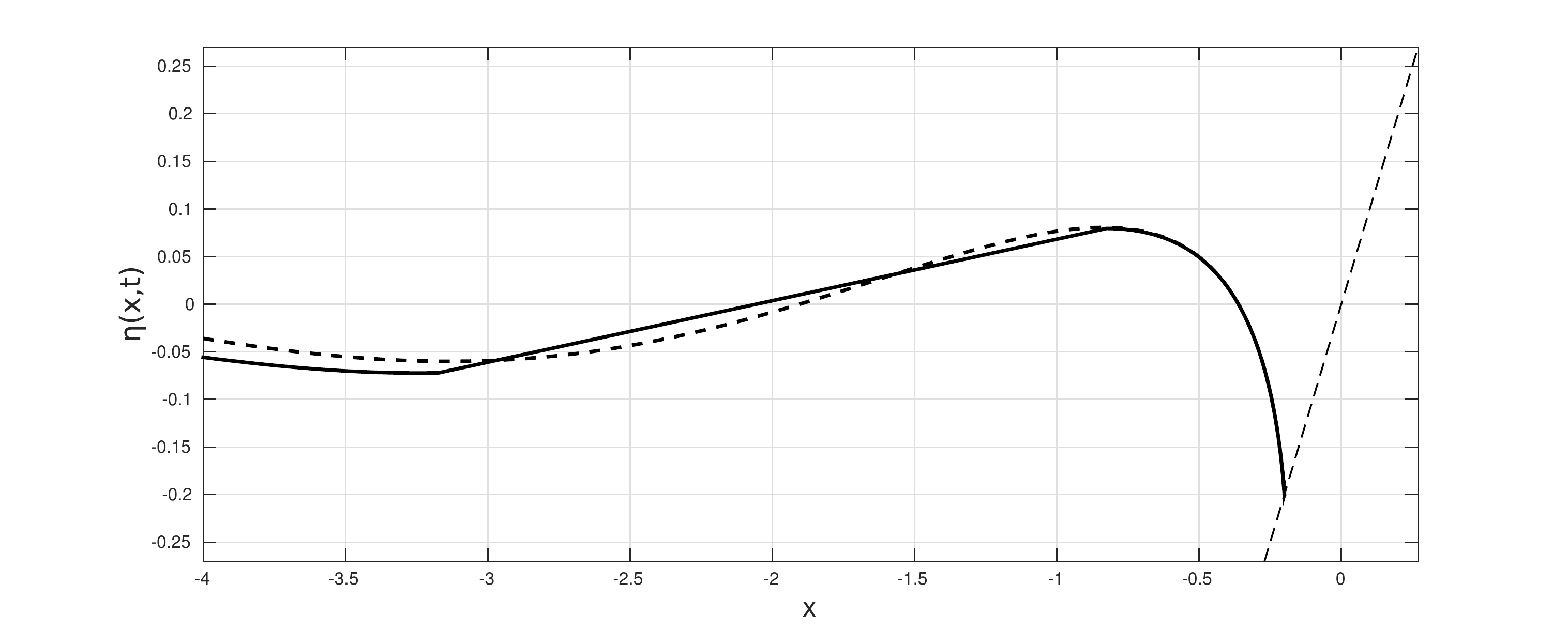}}\hfill%
\parbox{\LW}{$t_4=2.35$}\hfill \\
\caption{\footnotesize Free surface evolution with and without vorticity. 
The parameters are $A=0.2$, $\Gamma_0=0.0025$, $\alpha \Gamma_1=0.1$, $\omega = 1$.
The four plots are at time $t_1=0.89$, $t_2=1.38$, $t_3=1.87$ and $t_4=2.35$.
The nonzero vorticity has the effect of introducing a slight setdown on the left-hand side.}
\end{center}
\end{figure}

While these formulae give representations of solutions of 
(\ref{system1}) and (\ref{system2}), it is not completely
straightforward to understand these solutions in $(x,t)$-coordinates. 
Indeed, in order to plot these exact solutions in terms of $(x,t)$-coordinates, a numerical approach
is needed. A direct approach has been outlined for the problem without a shear flow \cite{AntuonoBrocchini3},
but it is unclear whether this method will work in the current situation with non-zero vorticity.
Therefore, let us briefly outline the numerical implementation.
First, expressions for $\phi$, $W$, $t$, $x$, $\eta$, and $u$, which are all functions of $(c,\lambda)$
are used to fill arrays of numbers as $c$ and $\lambda$ run through certain prescribed sets of values.
In order to plot the free surface elevation as $\eta=\eta(x,t)$, 
we use the two matrices for $t(c,\lambda)$ and $x(c,\lambda)$ as independent variables,
and tag the indices corresponding to certain values of $x$ and $t$ (to a prescribed tolerance).
Then, we use these same indices in the matrix for $\eta$ in order to find $\eta$ as a function
of $x$ and $t$. With this simple scheme, plots of $\eta(x,t)$ are possible.
The visualization of the horizontal fluid velocity $u=u(x,t)$ can be done in a similar way.
The solution is single-valued so long as the Jacobian $\frac{\partial(x,t)}{\partial(c,\lambda)}$ is nonzero. 
Therefore, the constants $A$, $\alpha$, $\Gamma_0$, $\Gamma_1$, $\omega$ 
and the arrays of $(c,\lambda)$ are all chosen so that a single-value solution is obtained.

Several plots are shown in Figures 3, 4 and 5. Figure 3 focuses on the comparison between
the solutions found here with small $\alpha \Gamma_1$ (solid curves) 
and solutions found using the method of Carrier and Greenspan
(dashed curves). It can be seen that the main effect of the background vorticity
is to induce a small setdown on the left-hand side (a minor downward deflection of the mean water level). 
Note also that the construction laid down here depends on non-zero $\Gamma_1$, so 
that the good agreement with the Carrier-Greenspan solutions validates our method.
On the other hand, Figures 4 and 5
focus on the comparison of different strengths of vorticity. Here, it can be seen that
while the run-up and run-down on the beach is identical, the amplitude of the wave
on the left-hand side is smaller in the case of larger vorticity. Note that in these
cases (as discussed in section 3), the parameter $\alpha \Gamma_1$ serves
to measure the combined effect of the strength of the slope and the vorticity,
since this parameter appears prominently in the non-dimensional version of the
equations.

\begin{figure}[h!]
\centering					
\includegraphics[width=10cm]{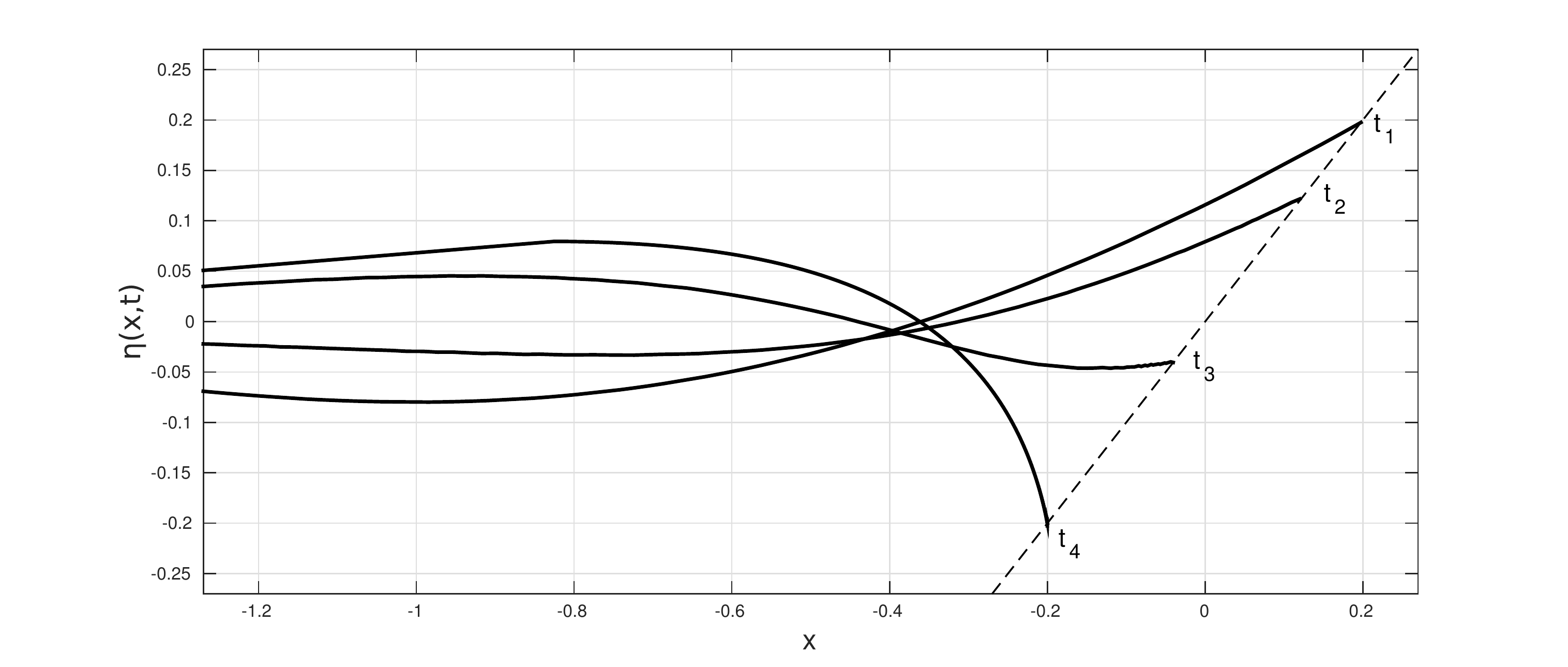}
\caption{\footnotesize Solution of a wave running up on a sloping beach with $\alpha \Gamma_1 = 0.1$. 
The solution parameters are $A=0.2$, $\Gamma_0 = 0.0025$ and $\omega = 1$. 
The solution is plotted at $t_1=0.89$, $t_2=1.38$, $t_3=1.87$, $t_4=2.35$.}
%\label{fig:slopebeachshear}
\end{figure}
\begin{figure}[h!]
\centering					
\includegraphics[width=10cm]{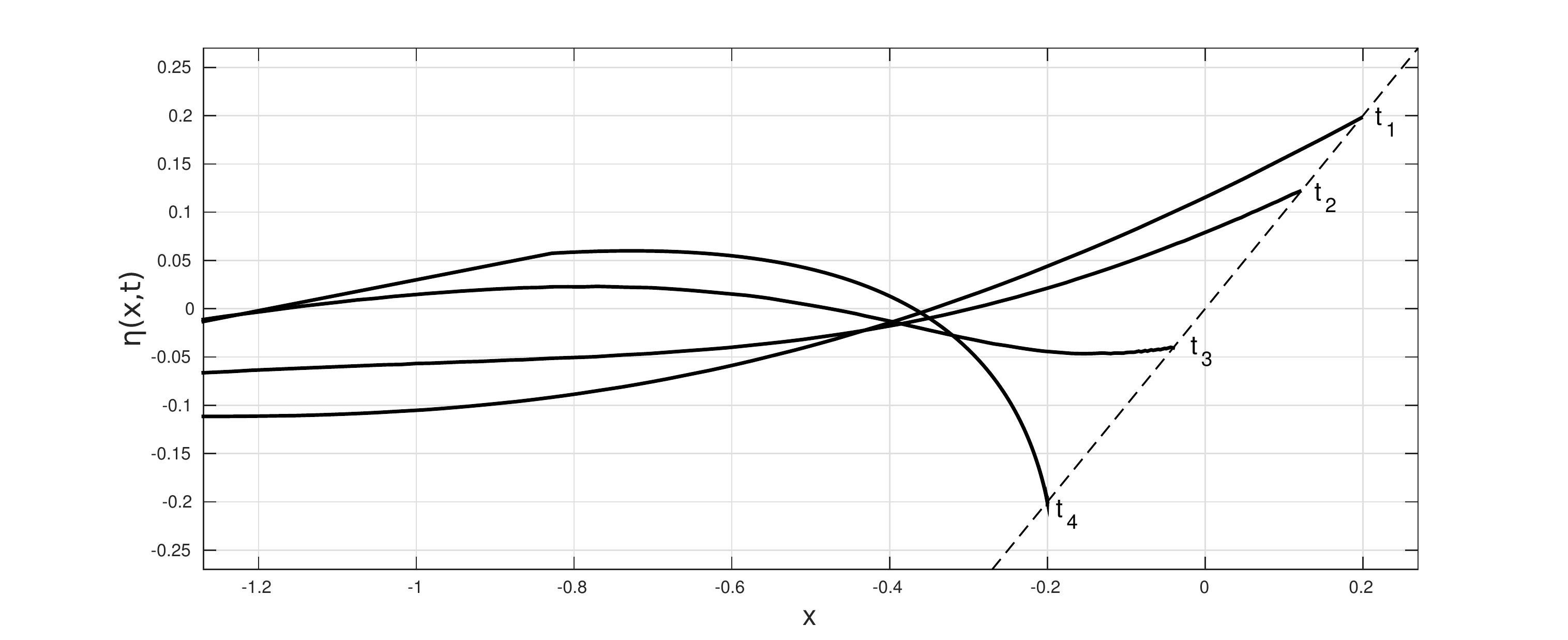}
\caption{\footnotesize Solution of a wave running up on a sloping beach with $\alpha \Gamma_1 = 0.5$. 
The solution parameters are $A=0.2$, $\Gamma_0 = 0.0025$ and $\omega = 1$. 
The solution is plotted at $t_1=0.89$, $t_2=1.38$, $t_3=1.87$, $t_4=2.35$.}
%\label{fig:slopebeachshear}
\end{figure}

\section{Appendix}
For the sake of completeness, the shallow-water equations with a background shear flow over a sloping beach 
will be derived.
This derivation complements other already existing asymptotic models
with background shear, such as presented in \cite{AK3,Choi,ThomasKharifManna}.
For a one dimensional flow we consider the velocity component to be $V(x,z,t)=U(z)+u(x,t)$,
where the linear shear current is given by $U(z)=\Gamma_0+\Gamma_1z$ with $\Gamma_0$ and $\Gamma_1$ being constants. 
For an incompressible and inviscid fluid, the equation for conservation of mass for
the control interval delimited by $x_1$ and $x_2$ on the $x$-axis is written as 
\begin{equation}\nonumber
\frac{d}{dt} \int_{x_{1}}^{x_{2}}  H(x,t) \:dx +\left[ \int_{-h(x)}^{\eta(x,t)} V(x,z,t) \:dz  \right]_{x_{1}}^{x_{2}} = 0,
\end{equation}
where $H(x,t)=\eta(x,t)+h(x)$ is the total depth. Integrating in $z$ yields
\begin{equation}\nonumber
 \int_{x_{1}}^{x_{2}} \eta_t+ \Big( \Gamma_0H+\frac{\Gamma_1}{2}(\eta^2-h^2)+uH \Big)_x \:dx   =0.
\end{equation}
Since $x_1$ and $x_2$ are arbitrary, the integrand must vanish identically, so that we get
the local mass balance equation
\begin{equation}\label{appendix,mass}
\eta_t+ \Big( \Gamma_0H+\frac{\Gamma_1}{2}(\eta^2-h^2)+uH \Big)_x=0.
\end{equation}
For later use, we can rewrite this equation in a slightly different form as
\begin{equation} \label{appendix,mass,differentway}
\eta_t+ \Big( \Gamma_0\eta+\frac{\Gamma_1}{2}\eta^2+u\eta \Big)_x  +
\Big( \Gamma_0h-\frac{\Gamma_1}{2}h^2+uh \Big)_x=0.
\end{equation}

Next, we will consider the momentum balance in the $x$-direction.
Recall that the only forces acting on the control volume are the pressure force,
and that the shallow-water approximation entails the assumption that the pressure is hydrostatic. 
The conservation of momentum is written as
\begin{equation}\label{appendix,momentum,integralform}
\frac{d}{dt}\int_{x_1}^{x_2}\int_{-h}^{\eta}V\:dxdz + \left[\int_{-h}^{\eta}V^2\:dz\right]_{x_{1}}^{x_{2}}
+ \left[\int_{-h}^{\eta} g(\eta-z)\:dz\right]_{x_{1}}^{x_{2}} = -\int_{x_1}^{x_2} \alpha g(\eta+h)\:dx,
\end{equation}
where the second term on the left is the momentum flux through the lateral boundaries
of the control volume at $x_1$ and $x_2$, and the third term on the left is the pressure
force on these lateral boundaries. The term on the rights represents the pressure force
in the negative $x$-direction due to the inclined bottom profile.
The integral in the second term can be calculated to be
\begin{align}\nonumber
\int_{-h}^{\eta}V^2\:dz 
%&=  \int_{-h}^{\eta}U^2+2Uu+u^2\:dz \\ \nonumber
%						&= \left.\left(\Gamma_0^2z+\Gamma_0\Gamma_1z^2+\frac{\Gamma_1^2}{3}z^3+2\Gamma_0uz+
%						   \Gamma_1uz^2+u^2z \right)\right|_{-h}^{\eta} \\ \nonumber
						&= \Gamma_0^2H+\Gamma_0\Gamma_1(\eta^2-h^2)+\frac{\Gamma_1^2}{3}(\eta^3+h^3) 										+2\Gamma_0uH+ \Gamma_1u(\eta^2-h^2)+u^2H.   \nonumber
\end{align} 
Again, the integrand must vanish pointwise
so eq.(\ref{appendix,momentum,integralform}) requires that
\begin{multline} \nonumber
\Big( \Gamma_0H+\frac{\Gamma_1}{2}(\eta^2-h^2)+uH \Big)_t + \\
  \Big( \Gamma_0^2H+\Gamma_0\Gamma_1(\eta^2-h^2) +\frac{\Gamma_1^2}{3}(\eta^3+h^3)+2\Gamma_0uH+ \Gamma_1u(\eta^2-h^2)+u^2H +\frac{g}{2}H^2 \Big)_x  %\\
  = -\alpha gH.
 \end{multline}
This equation can be simplified significantly by combining it with eq.(\ref{appendix,mass}). 
Removing terms of the form ($\Gamma_0\: \cdot$ eq.(\ref{appendix,mass})) 
and ($u\: \cdot$ eq.(\ref{appendix,mass})), the equation becomes
\begin{multline} \nonumber
\Gamma_1\left[\eta\eta_t +\frac{1}{2}\Gamma_0\Gamma_1(\eta^2-h^2)_x+\frac{1}{3}\Gamma_1^2(\eta^3+h^3)_x+ \Gamma_1u_x(\eta^2-h^2)+\frac{1}{2}\Gamma_1u(\eta^2-h^2)_x\right] \\
+ H\left(u_t+\Gamma_0u_x+uu_x+gH_x+\alpha g\right) = 0.
 \end{multline}
After some algebra, the equation can be written as
\begin{multline}\nonumber
\Gamma_1 \left[\eta \Big( \eta_t + \Big( \Gamma_0\eta+\frac{\Gamma_1}{2}\eta^2+u\eta \Big)_x\Big)- 
h\Big(\Gamma_0h-\frac{\Gamma_1}{2}h^2+uh\Big)_x \right] \\
+ H\Big( u_t+\Gamma_0u_x+uu_x+gH_x+\alpha g \Big) = 0.
 \end{multline}
By eq (\ref{appendix,mass,differentway}), we obtain
\begin{multline}\nonumber
\Gamma_1 \left[\eta\Big( -\Big(\Gamma_0h-\frac{\Gamma_1}{2}h^2+uh \Big)_x \Big)- 
h\Big( \Gamma_0h-\frac{\Gamma_1}{2}h^2+uh \Big)_x \right] \\
+ H\Big( u_t+\Gamma_0u_x+uu_x+gH_x+\alpha g \Big) = 0,
\end{multline}
which is the same as
\begin{equation}\nonumber
\Gamma_1 H \Big(-\Gamma_0h+\frac{\Gamma_1}{2}h^2-uh\Big)_x 
+ H\Big( u_t+\Gamma_0u_x+uu_x+gH_x+\alpha g \Big) = 0.
\end{equation}
Excluding $H(x,t)$ from the equation and by inserting the undisturbed water depth denoted by $h(x)=-\alpha x$, gives us
\begin{equation}\nonumber
\alpha\Gamma_0\Gamma_1+\alpha^2\Gamma_1^2x+\alpha\Gamma_1xu_x+\alpha\Gamma_1u+u_t+\Gamma_0u_x+uu_x+g\eta_x=0.
\end{equation}
Defining the function $s(x)=\Gamma_0+\alpha\Gamma_1x$, leads to the equation
\begin{equation}\nonumber
ss_x+u_xs+uu_x+us_x+g\eta_x+u_t=0,
\end{equation}
which can be written as
\begin{equation*}
u_t+ \Big( us+\frac{s^2}{2}+\frac{u^2}{2}+g\eta \Big)_x = 0. \\
\end{equation*}
\vskip 0.3in

\noindent
{\bf Acknowledgments:}
{\small This research was supported in part by the Research Council of Norway
under grants 213474/F20 and 239033/F20.
}

\end{document}